# Bicontinuous interfacially jammed emulsion gels with nearly uniform sub-micrometer domains via regulated co-solvent removal

*Tiancheng Wang, Robert A. Riggleman, Daeyeon Lee\* and Kathleen J. Stebe\**

Department of Chemical and Biomolecular Engineering, University of Pennsylvania, Philadelphia, PA 19104, United States
E-mail: daeyeon@seas.upenn.edu, kstebe@seas.upenn.edu

Abstract

Porous materials possess numerous useful functions because of their high surface area and ability to modulate the transport of heat, mass, fluids, and electromagnetic waves. Unlike highly ordered structures, disordered porous structures offer the advantages of ease of fabrication and high fault tolerance. Bicontinuous interfacially jammed emulsion gels (bijels) are kinetically trapped disordered biphasic materials that can be converted to porous materials with tunable features. Current methods of bijel fabrication result in domains that are micrometers or larger, and non-uniform in size, limiting their surface area, mechanical strength, and interaction with electromagnetic waves. In this work, scalable synthesis of bijels with uniform and sub-micrometer domains is achieved via a two-step solvent removal process. The resulting bijels are characterized quantitatively to verify the uniformity and sub-micrometer scale of the domains. Moreover, these bijels have structures that resemble the microstructure of the scale of the white beetle Cyphochilus. We find that such bijel films with relatively small thicknesses (< 150 μm) exhibit strong solar reflectance as well as high brightness and whiteness in the visible range. Considering their scalability in manufacturing, we believe that VIPS-STRIPS bijels have great potential in large-scale applications including passive cooling, solar cells, and light emitting diodes (LEDs).

Porous materials designed with spatially distributed features are extensively used in diverse applications including optics, chemical reaction, filtration, and structural materials. In comparison to their highly ordered counterparts, disordered porous structures have the advantage of being fault tolerant, and their relative ease of fabrication reduces manufacturing costs. Bicontinuous interfacially jammed emulsion gels (bijels) are multiphasic colloidal gels with two interconnected immiscible liquid phases stabilized by jammed nanoparticles at the liquid-liquid interface.[1,2] Upon solidification of one phase, a porous structure with continuous pores can be obtained. Taking advantage of their unique spinodal bicontinuous morphology, bijel templated materials have been employed in diverse applications including micro-reactors[3], separation membranes[4], biomedical implants[5], and energy storage devices[6–9]. Although many different paths of fabrication have been proposed, limited control over bijel morphology and the complexity of bijel fabrication have restricted their vast potential, for example, as optical devices and structural materials.

Typically, bijel formation relies on the kinetic arrest of spinodal decomposition of two immiscible phases triggered by chemical or temperature changes. During the phase separation process, nanoparticles become irreversibly trapped at the fluid-fluid interface; if the nanoparticles wet both phases, the energy required to remove them from the interface can be several orders of magnitude higher than thermal perturbation[10]. As the interface coarsens, these interfacially-trapped particles form a jammed layer that arrests the coarsening of the intertwined fluid channels. Bijels were initially formed via a thermal quenching of a binary mixture to initiate spinodal decomposition[1,11], spurring research into their application[12]. However, the interfacial tension between the two phases that develop during spinodal decomposition initiated by thermal quenching is not sufficiently high enough to effectively trap nanoparticles that are smaller than a few tens of nanometers[13]. For these thermally quenched bijels, larger particles (> 200 nm), which have higher interfacial trapping energies, are typically used to stabilize the structure, resulting in typical domain sizes on the scale of 10 μm[1,14].

Recently, methods that rely on the removal of co-solvent from a ternary mixture to induce phase separation have been developed[4,15,16]. In comparison to thermal quench-based methods, chemical quenching allows access to much higher interfacial tensions as phase separation proceeds[17]. However, the flow generated by co-solvent removal has a significant impact on the resulting bijels' structures. In solvent transfer-induced phase separation (STRIPS), the partitioning of co-solvent to the outer liquid phase induces a strong flow that drives the formation of anisotropic domains aligned in the direction of the co-solvent flux. In addition, STRIPS bijels have heterogeneous domain sizes; STRIPS bijels have sub-micrometer domains on their outermost surfaces, and larger, coarsened internal structures[15]. STRIPS bijels with such aligned asymmetric structures have shown superior performance as filtration membranes[4]. We have also introduced the vaporization induced phase separation (VIPS) method[20] in which the gentle evaporation of a co-solvent minimized the influence of fluid flow on structure formation. However, the rate of co-solvent removal in VIPS is not high enough to form internal domains with sub-micrometer features, even with the help of airflow accelerating the evaporation. For some applications it would be critical to preserve the 3D nanostructure formed by the phase separating fluids with minimal influence of parasitic forces so that the nature and full potential of the spinodal structure can be studied and harnessed.

The ability to form bijels with uniform, sub-micrometer domains would enable their utilization in reactive separation, fluid-fluid extraction, flow batteries, fabrication of anti-cracking composite matrix[28] as well as optics and photonics. For liquid phase reaction and extraction, the large surface area facilitates reaction and mass transfer between two phases[17,21,22] whereas the uniform channels provide homogeneous flow/solute distributions[23–25] essential, for example, to flow battery function. For photonics, recent simulation results have proposed that spinodal structures exhibit hyperuniformity[26] making bijel-templated materials potentially useful for designing photonic band gaps[27]. Furthermore, some porous biomaterials with spinodal-like structures have unique optical responses that act as amorphous photonic crystals[29–31]. Bijels with sub-micrometer features have recently been prepared by using an organic solvent, toluene, as the outer oil phase and 1-propanol as co-solvent to further increase interfacial tension between the two immiscible phases during phase separation[17]. This method, nevertheless, produces bijels with directionally-aligned water macro-voids and requires an organic continuous phase, making it challenging to produce large area films and membranes for various applications.

Here, we describe the successful fabrication of bijels with isotropic, nearly uniform and sub-micrometer domains via a process that take advantage of sequential VIPS and STRIPS. In this VIPS-STRIPS method, the formation of a top bijel layer is induced by co-solvent evaporation into the vapor phase via VIPS is followed by removal of co-solvent to an external aqueous phase via STRIPS. The initial VIPS bijel outer layer, formed in air, regulates the partitioning of co-solvent to the outer aqueous phase during the subsequent STRIPS ripening of the bijel. In the following, we examine the final structures with microscopic imaging and fast Fourier transform (FFT) analysis. Further, we demonstrate that 130 μm thick bijel films with uniform submicrometer domains exhibit excellent broadband reflection over the solar spectrum via multiple scattering, producing high solar reflectance. Our results suggests that bijels with isotropically distributed, nearly uniform sub-micrometer domains could potentially be used as optical/photonic elements to manipulate the propagation and transport of electromagnetic waves.

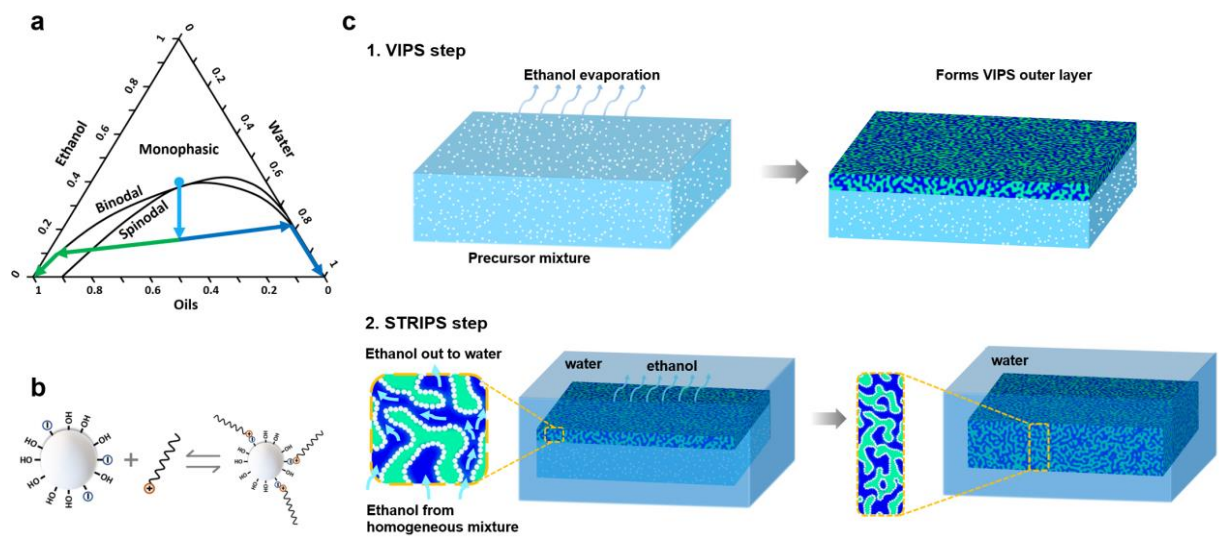

**Figure 1. VIPS-STRIPS bijel film formation. a)** Representative ternary phase diagram as a function of volume fractions of oil, water and ethanol. A homogeneous precursor mixture is quenched into the biphasic region upon removal of ethanol. The quenching path is indicated by the light blue arrow. **b)** Silica nanoparticle modification with CTAB. **c)** Bijel film formation via an initial VIPs step in which a bijel outer layer is formed followed by a STRIPS step in which the outer layer regultes the STRIPS co-solvent removal process.

The precursor solution used to prepare bijels is a ternary mixture composed of two immiscible liquids, such as water and oil, which upon the addition of the co-solvent, such as ethanol, become miscible to form a homogeneous single phase. To this mixture, surfactants and nanoparticles are added; the surfactants modify the wetting property of the nanoparticles *in situ* during phase separation. The phase behavior of the mixture is shown schematically by the ternary phase diagram shown schematically in Figure 1a. Above the binodal line, the mixture contains enough ethanol to induce mixing so that the mixture is in the monophasic region. As ethanol is removed, the composition of the mixture crosses the binodal and spinodal lines as indicated by the blue arrow, and triggers phase separation. To induce spinodal decomposition, the quenching path should be close to the critical point at the intersection of the spinodal and binodal lines. During spinodal decomposition, two interconnected continuous phases form with large interfacial areas. Absent nanoparticles, the structure would coarsen to reduce interfacial area until the two phases are completely separated. However, with the formation of a dense layer of nanoparticles jammed at the interface of the two liquid phases, the self-assembled structure of bicontinuous interfacially jammed emulsion gels (bijels) can be kinetically arrested.

It is crucial for the particles to neutrally wet the two phases to irreversibly trap them at the interface and to prevent the formation of dispersed droplets. Here we use silica nanoparticles (22 nm) combined with cetyltrimethyl ammonium bromide (CTAB) to adjust their wettability via electrostatic interaction[32,33], as illustrated in Figure 1b.

We illustrate in Figure 1c the fabrication process of uniform sub-micrometer domain bijel films via the VIPS-STRIPS process with a ternary system composed of water and 1,6-hexanediol diacrylate (HDA) as immiscible liquids and ethanol as co-solvent. In the VIPS step of the proccess, the homogeneous precursor mixture is applied to a substrate to form a film under ambient conditions. Within a few seconds, a thin layer of VIPS bijel is formed by the evaporation of ethanol, accelerated by an air stream (5 m/s) parallel to the evaporating surface[20]. Subsequently, in the STRIPS stage of the process, the entire film is submerged into water to initiate further bijel maturation via STRIPS. During this stage, the ethanol content in the remaining homogeneous mixture must flux through the VIPS outer layer in order to partition to the outer aqueous phase. This thin outer layer of gelled bijel eliminates the strong flow of ethanol within the phase separating region by regulating the ethanol flux through the tortuous channels. Furthermore, the rigidity of the VIPS outer layer protects the film from strong disturbances, avoiding the formation of defects like wrinkles and holes that can often form in relatively thick (>100 μm) bijel films on solid supports when STRIPS is implemented without the inital VIPS step. After extensive removal of ethanol, the bijel structure develops throughout the entire film. With the protection of the top VIPS outer layer, defect-free bijels of various thickness from 10 to 150 μm can be prepared, as shown in Figure S1. Photo-initiator, added to the mixture, partitions to the HDA phase, allowing the oil phase to be polymerized via a 10s UV-irradiation to form a bicontinuous porous polymer film. This bijel film spontaneously detaches from the glass substrate upon solidification, forming a free-standing bijel film.

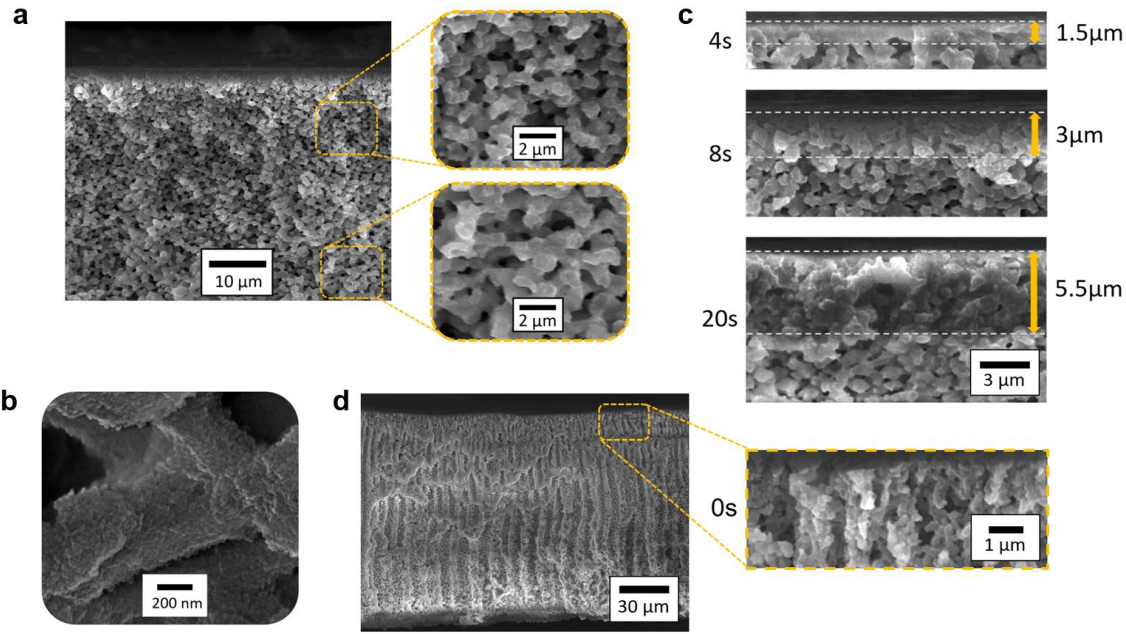

**Figure 2.** Cross-sectional scanning electron microscopy (SEM) micrographs of polymerized bijel films. A) VIPs-STRIPS bijel film. B) Jammed silica nanoparticle layer on the interface. C) The development of VIPS outer layer from the surface to the internal region with different evaporation time. D) Polymerized STRIPS bijel film from the same precursor mixture without VIPS treatment.

The polymerization process not only solidifies the oil phase, but also locks in the nanoparticle layer at the surface of the bijel structure. Such polymerized bijel films can be visualized under a scanning electron microscope (SEM) after drying the water phase (Figure 2a-d). In cross-sectional SEM images, the size and size distribution of the polymer phase and pore phase (previously filled with water) are evident. The domains in these bijels are sub-micrometer as seen in Figure 1a, although the exact feature size is difficult to determine by eye for such a spinodal structure. Furthermore, by comparing features at different depths from the surface as shown in the two zoom-in insets of Figure 1a, we see only minor differences in their domain sizes which we characterize more in detail below. If we further zoom in (Figure 2b), jammed nanoparticle layers at the polymer surface can be visualized, proving the kinetically arrested nature of bijel formation mechanism. Moreover, when looking at the cross-sectional image (Figure 2a), there is no obvious alignment of features along the vertical direction. Instead, the distribution and orientation of bijel domain appears to be isotropic throughout the entire film. We attribute the ability to form this structure to the protective VIPS outer layer, which successfully slows the flow of ethanol and prevents the formation of anisotropic features.

For comparison, the same precursor mixture is used to fabricate a bijel via STRIPS, without the VIPS pre-treatment. Inspection of the cross-sectional image of the resulting bijel (Figure 2d) clearly reveals the effects of rapid ethanol fluxes towards the upper surface; the domains of the STRIPS bijel film are clearly aligned in the same direction of such flow, even though the presence of 10 wt% ethanol in outer aqueous phase retards the mass flux. Aside from eliminating the alignment of features, the VIPS pre-treatment also eliminates the gradient of domain sizes along the bijel thickness. Without the VIPS pre-treatment, the initial rapid partitioning of ethanol induces deep quenching near the surface. This deep quench facilitates nanoparticle capture and leads to the rapid arrest of small domains sizes (< 200 nm) in this region, as shown in Figure 2d inset. These small domains further hinder mass transfer, leading to large, highly coarsened internal domains. Unlike STRIPS, VIPS relies on the relatively slow flux of ethanol to induce phase separation which allows formation of > 500 nm domains in the surface region. We believe that this VIPS outer layer regulates ethanol flow and therefore the size of the internal structures. Since VIPS and STRIPS follow slightly different quenching paths, primarily due to the change of water concentration as illustrated in Figure S2, the VIPS outer layer is richer in oil than the region formed via STRIPS. A range of VIPS surface structures are observed in the cross-sectional images of bijel films as the time of VIPS treatment is varied (Figure 2c). We find that the VIPS layer developed in 4 s with a thickness of around 1.5μm is able to resist the rapid flow of ethanol. By increasing evaporation time from 4 s to 8 s, the thickness of the VIPS surface layer increases from around 1.5 μm to 3 μm. When the VIPS treatment time reaches 20 s, its thickness approaches 5.5 μm and the top VIPS outer layer starts to lose its original structure due to the substantial loss of water during evaporation.

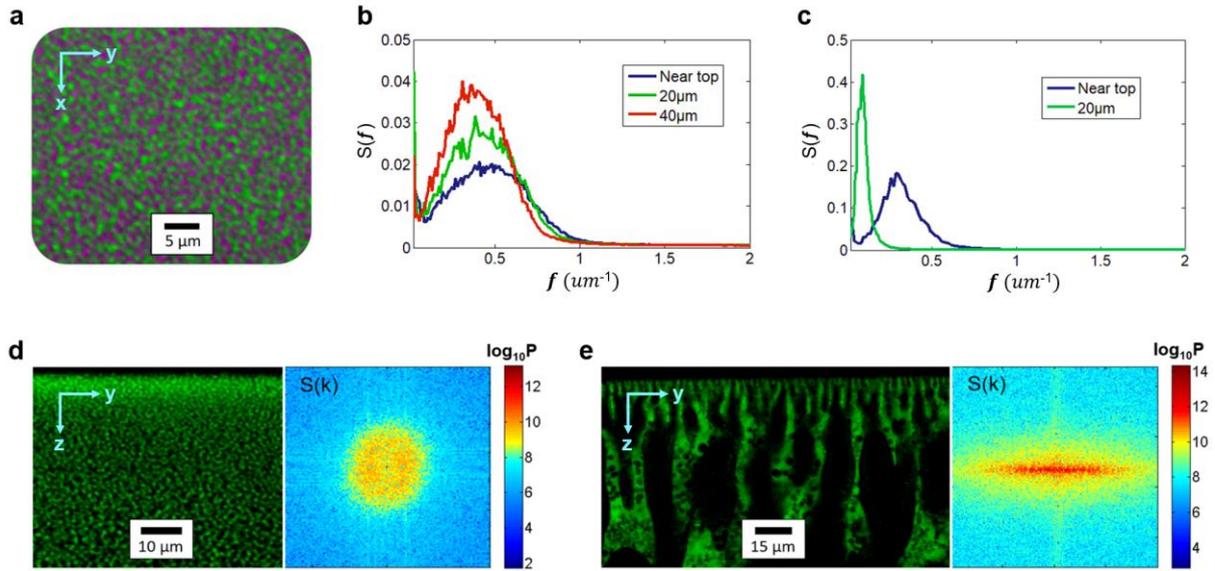

**Figure 3. Quantitative analysis of confocal micrographs of VIPS-STRIPS bijel film. The green domains are the polymerized HDA phase labeled with Nile Red and the dark/purple domains are the void phases filled with diethyl phthalate (DEP). a) The exactly complementary distribution of the polymer and DEP domains with different fluorescence. b) Radially averaged power spectra of VIPS-STRIPS bijel structures at different vertical distance from the top surface. c) Radially averaged power spectra of a STRIPS bijel at different vertical distance from the top surface. d-e) Cross-sectional confocal fluorescent images of VIPS-STRIPS and STRIPS bijels and the corresponding 2D Fourier power spectra.**

To confirm the bicontinuity and uniformity of the structures formed by the sequential VIPS and STRIPS treatment, we use confocal microscopy. A dye permeation experiment is conducted in the middle of the film using the same protocol as previously reported[15]. Briefly, a bijel film with polymerized oil phase (green) is immersed in dyed diethyl phthalate (magenta) phase. The structural integrity of the polymer phase indicates that the oil phase is continuous, whereas the permeation of dye (magenta) to all the void space confirms the continuity of the water phase.

To better characterize the internal structure of these VIPS-STRIPS bijels, we take advantage of Fast Furrier Transform (FFT) analysis of confocal microscopy images. The result of the 2D FFT of the image, $I(\boldsymbol{k})$, is then converted to its power spectrum $S(\boldsymbol{k}) = I(\boldsymbol{k}) \cdot \overline{I(\boldsymbol{k})}$, and the 1D profiles shown in Figures 3b and c are radially averaged over orientations of **k**, and $f = |\boldsymbol{k}|/2\pi$. Traditionally, the feature size of a bijel is defined as the width of one domain which is half of the period length, assuming the two phases are of similar sizes. We can estimate a characteristic feature size using $d = \frac{1}{2f^*}$, where the star indicates the value of $f$ associated with the maximum of $S(f)$ and the factor of 2 converts the wavelength to feature size. From Figure 3b, we find an apparent $d \approx 1\ \mu m$, and this value is slightly larger than the sub-micron estimates from the SEM images, possibly due to the analysis of a 2D Fourier series on a 3D structure. As shown in Figure 3b, the power spectra of a VIPS-regulated STRIPS bijel at different vertical distances from the uppermost surface show peaks that remain at approximately the same location albeit with a small shift toward the lower wavenumber; these spectra indicate that the increase in the domain size through the vertical direction is less than 25% over a distance of around 20 repeat units of a single domain, equivalent to 40 μm (a log-log scale plot can be found in SI). For comparison, the feature size of a traditional STRIPS bijel increases by 250% over a vertical distance of 20 μm, as shown in Figure 3c. Moreover, FFT of the z-x plane confocal image shows that the bijel film fabricated without VIPS treatment shows a strong intensity along the x axis, indicating the alignment of domains along the y axis due to rapid ethanol flow (Figure 3d). In contrast, the spectrum of VIPS-regulated STRIPS bijel is much more uniform, confirming the more isotropic nature of the structure (Figure 3e).

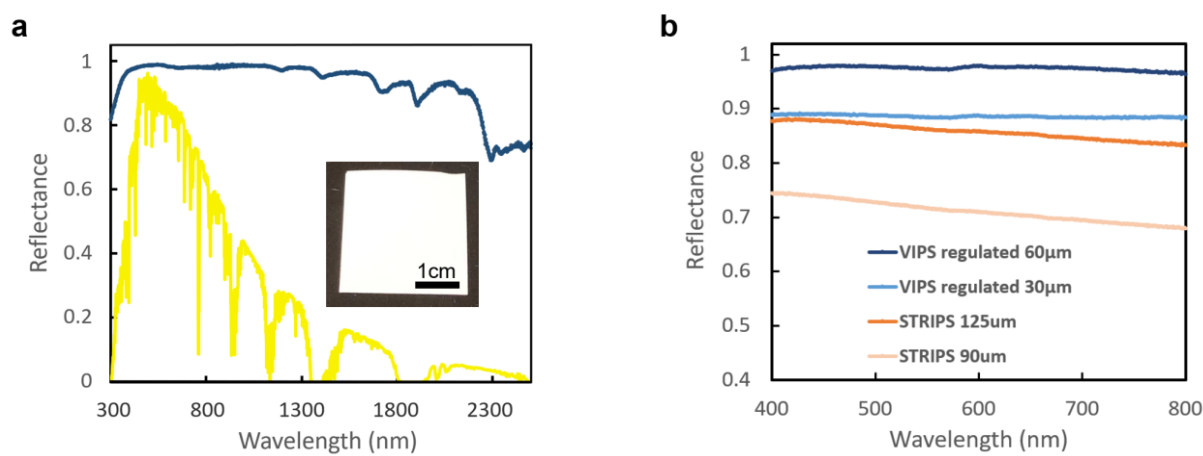

**Figure 4. Spectral reflectance characterization. a) VIPS regulated bijels of ~ 130 μm thickness. Yellow spectrum indicates normalized ASTM G173 Global solar spectrum. Inset shows macroscopic morphology of a ~ 130 μm thick VIPS-STRIPS film. After polymerization, the dried sample forms a thick white opaque free-standing film. b) Regulated/non-regulated bijel films of different film thickness with incident angle of 45°.**

Studies of the optical responses of disordered structures have led to novel findings such as the photonic response of hyperuniform structures[34,35] and the strong brightness and whiteness of beetle scale structures.[29,36–38] Many of these structures are based on or are quite similar to spinodal structures. In particular, the scale of a white beetle, Cyphochilus, comprises bicontinuous disordered structures that bear a striking resemblance to the structure of the VIPS-STRIPS bijel. The bicontinuous structure of the white beetle's scale leads to efficient scattering light to produce high whiteness despite their small thicknesses (8-15 μm). The realization of bijels with sub-micrometer domains with high uniformity makes them an interesting system to probe their response in the visible and near IR spectra.

As shown in Figure 4a, a free-standing VIPS-STRIPS bijel films with 130 μm thickness exhibits bright white appearance as well as effective broad-band reflection over the entire solar spectrum. According to the normalized ASTM G173 Global solar spectrum, 97% of the solar irradiance is reflected which rivals the high performance low-index porous solar reflectors reported in the literature[39–41]. Furthermore, a fairly thin VIPS-STRIPS bijel film with the thickness of around 30 μm exhibits ultra-high brightness and whiteness (400-800 nm) as shown in Figure 4b. In comparison, much thicker STRIPS-only bijel films exhibit lower brightness, and a substantial decrease in reflectivity with increasing wavelength. In general, the brightness and reflectance of porous media originate from the multiple scattering events determined by the morphology and spatial arrangement of scattering elements.

In summary, we have shown that bijels with sub-micrometer domains with nearly uniform domain sizes can be fabricated via sequential VIPS and STRIPS. The VIPS outer layer facilitates steady co-solvent removal to prevent the directional alignment of domains and drastically enhances the uniformity of domain sizes along co-solvent removal paths. The resulting bijels are characterized quantitatively to verify the effectiveness of the VIPS outer layer in producing bijels with nearly uniform sub-micrometer domains. Owing to their uniform nanostructures derived from spinodal decomposition, VIPS-STRIPS bijel films with relatively small thicknesses (< 150 μm) exhibit strong solar reflectance as well as high brightness and whiteness in the visible range. Considering the self-assembled nature and ease of fabrication, we believe that VIPS-STRIPS bijels have tremendous potential in large-scale applications including passive cooling, solar cells, and light emitting diodes (LEDs).

**Methods**

*The ternary liquid mixture is prepared with*: (1) 1,6-hexanediol diacrylate (HDA), (2) pH 3 water (adjusted by 1M HCl), (3) pure ethanol, and three additional components: (4) 50$_{wt}$% silica nanoparticle suspension (concentrated Ludox TMA in water, 22 nm, pH 3), (5) 200mM cetyl trimethyl ammonium bromide (CTAB) solution in ethanol, (6) 2-hydroxy-2-methylpropiophenone (HMP) as photo-initiator. As an example, these components are mixed in the following sequence to prepare a precursor mixture. 2.3g (1), 0.72g (3), 2.22g (5), 0.1g (2), 4.67g (4) and 0.05g (6). The mixture is shaken or vortexed until it becomes transparent. Subsequently, the sample is kept in dark to prevent undesired polymerization induced by ambient light. To enable fluorescent imaging of the oil phase, trace amount of Nile red can be added in the mixture that partitions to the oil phase upon phase separation.

*VIPS-STRIPS film formation:* The precursor mixture is placed on glass slides and blade-coated with an adjustable gap at room temperature. Meanwhile, an air pipe with a control valve is turned on to create airflow of different velocities ranging from 0 m/s to 15 m/s. After 4-20 s, the glass slide is immersed into a water bath (pH 3) for 2 mins. The sample is then irradiated with ultraviolet (UV) light (340 nm, 25 Wm$^{-2}$) underwater to polymerize the HDA phase within 1mins. The polymerized sample on a glass slide is placed in ethanol for 10 mins to make a freestanding film.

*Dye permeation experiment:* Polymerized bijel film samples are immersed in DEP dyed with 9,10-bis(phenylethynyl)anthracene (BPA). After 30 mins, a confocal microscope is used to monitor the spreading of DEP in the channels between polymer phases. The excitation wavelength is set to be 488 nm. Samples are imaged using fluorescence channels adjusted for the two dyes: 500-530 nm for the dyed DEP phase (BPA) and 625-725 nm for the polymer phase (Nile red).

*Spectral reflectance characterization:* Spectral reflectance of VIPS-STRIPS film samples are measured using Cary 5000 UV-Vis-NIR spectrophotometer equipped with a Praying mantis diffuse reflectance accessory. Optical-grade spectralon was used as the reference white material. The solar refelctance of the sample is defined by the equation:

$$\bar{R}_{solar}(\theta) = \frac{\int_0^\infty I_{solar}(\lambda) \cdot R_{solar}(\theta,\lambda) d\lambda}{\int_0^\infty I_{solar}(\lambda) d\lambda}$$

where θ is the angle of incidence from the surface normal, λ is the wavelength, $I_{solar}(\lambda)$ is the ASTM G173 Global solar intensity spectrum, and $R_{solar}(\theta, \lambda)$ is the surface's angular spectral reflectance. In this study, θ is fixed at 45°.

## ACKNOWLEDGMENT

The authors acknowledge financial support from Penn MRSEC through NSF (NSF DMR 1720530) and from NSF CBET 1945841. The authors thank Prof. Cherie R Kagan for access to spectrometers. This work was performed in part at the Singh Center for Nanotechnology at the University of Pennsylvania, a member of the National Nanotechnology Coordinated Infrastructure (NNCI) network, which is supported by the National Science Foundation (Grant No. NNCI-1542153). The electron microscopy facilities are supported by NSF through the University of Pennsylvania Materials Research Science and Engineering Center (MRSEC) (DMR-1720530).

Supporting Information

**Bicontinuous interfacially jammed emulsion gels with nearly uniform sub-micrometer domains via regulated co-solvent removal**

*Tiancheng Wang, Robert A. Riggleman, Daeyeon Lee[*] and Kathleen J. Stebe[*]*

**Control over VIPS-STRIPS bijels film thickness**

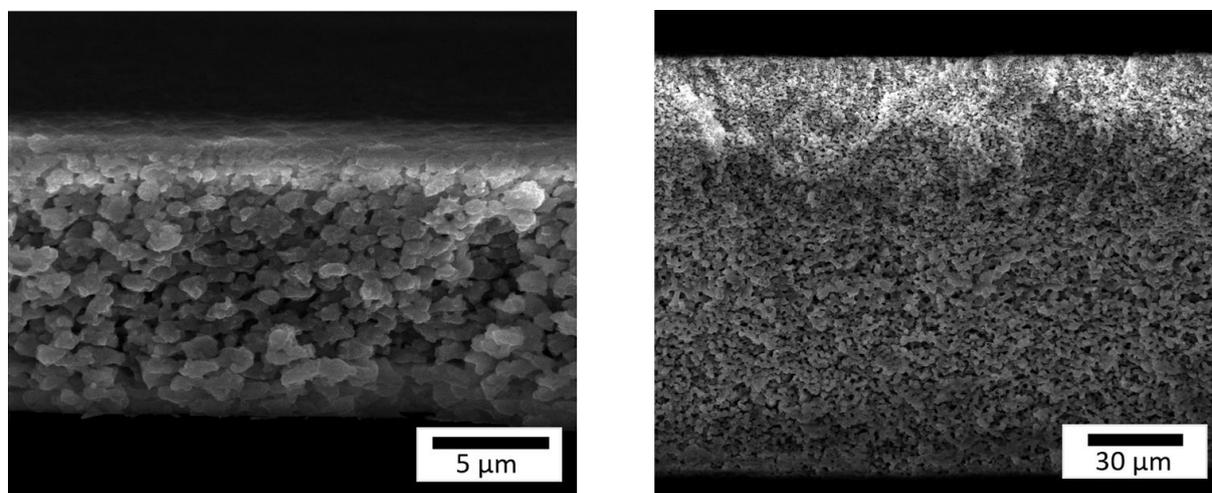

**Figure S1.** Cross-sectional SEM images of VIPS-STRIPS bijels film: (a) VIPS-STRIPS bijels with film thickness of around 10 μm. (b) VIPS-STRIPS bijels with film thickness over 130 μm.

In the VIPS-STRIPS method, as shown in Figure S1, control over the film thickness is achieved by varying the thickness of the precursor mixture film. For example, when applied by blade coating, the thickness of the precursor mixture film can be adjusted by using different doctor blades. The rigidity of the gelled VIPS outer layer protects the internal liquid layer from strong disturbances when entering outer aqueous phase, avoiding the formation of defects like wrinkles and holes that can often form in relatively thick (>100 μm) bijel films on solid supports.

**Different quenching paths of VIPS and STRIPS methods**

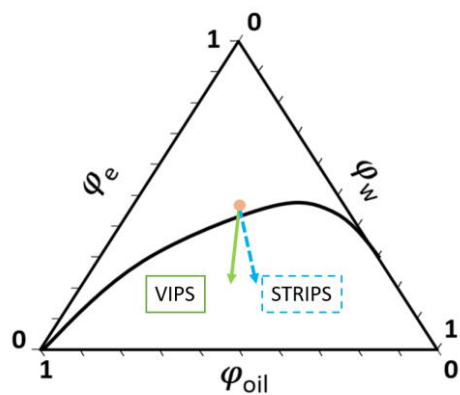

**Figure S2.** Representative quenching paths (removal of ethanol from precursor mixture) of VIPS and STRIPS methods are indicated by the green solid arrow and blue dashed arrow respectively.

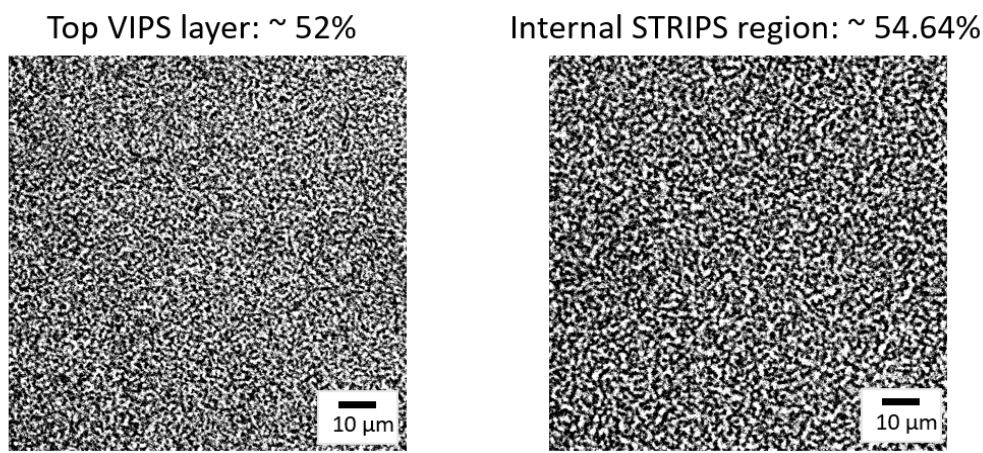

**Figure S3.** Binarized fluorescent images of VIPS-STRIPS top VIPS layer and internal STRIPS region and their water phase fractions. Dark phase represents water phase.

VIPS and STRIPS follow slightly different quenching paths as shown in Figure S2. In STRIPS method, water permeates into the phase separating mixture from the outer aqueous phase. Therefore, the STRIPS bijels becomes water rich. In contrast, in VIPS method, water co-evaporates with ethanol from the precursor mixture. As a result, the VIPS bijels becomes richer in oil. By analyzing the binarized fluorescent images of structures in the two regions, the fraction of water and oil phases can be calculated using ImageJ as shown in Figure S3. For a sample with a 8s VIPS treatment, the internal STRIPS region is richer in water phase by 2.64%.

**Fast Fourier transform analysis of VIPS-STRIPS bijel structures at different vertical distance from the top surface (in log-log scale)**

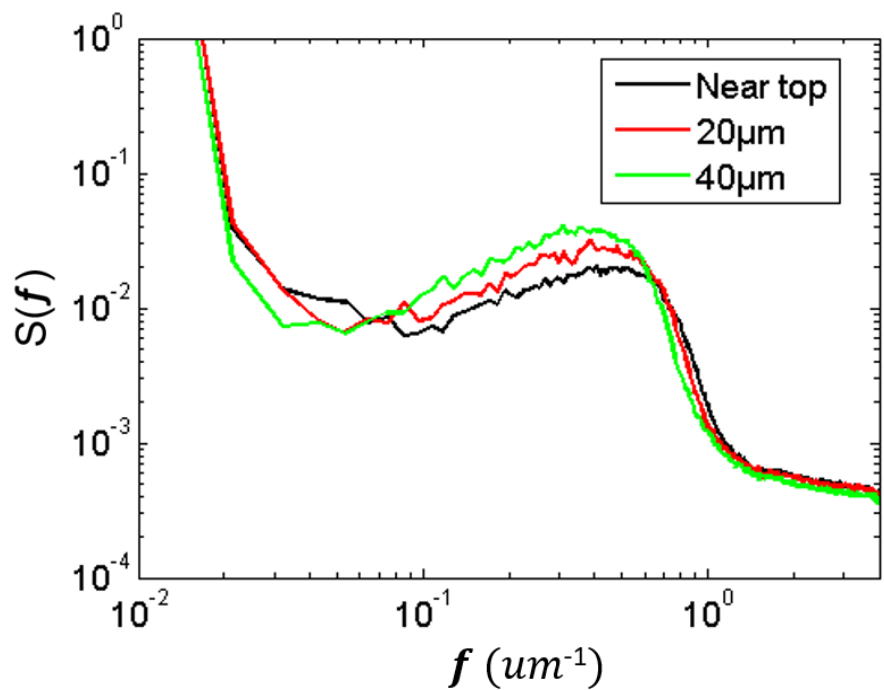

**Figure S4.** Radially averaged power spectra of VIPS-STRIPS bijel structures at different vertical distance from the top surface in log-log scale.